\documentclass[journal=jacsat,manuscript=article]{achemso} % uncomment for ACS format
\usepackage[version=3]{mhchem}
\usepackage{graphicx}
\usepackage{amsmath}
\usepackage{float}
\usepackage{amssymb}
\usepackage{placeins}
\usepackage{array}
\usepackage{subfig}
\usepackage{threeparttable}
\usepackage{color}
\usepackage[symbol]{footmisc}
\usepackage{physics}
\usepackage{makecell}
\usepackage{siunitx}
\usepackage{xr}
\usepackage{xstring}
\usepackage{multirow}
\usepackage{mathtools}
\usepackage{soul} % for strikeout text
\usepackage{changes}

\definecolor{tcolor}{rgb}{0, 0.5, 0.2}

\def\vac{\text{V}}
\def\vb{\vac_\text{B}}
\def\vn{\vac_\text{N}}
\def\nbvn{\text{N}_\text{B}\vac_\text{N}}
\def\gwbse{\text{G}_0\text{W}_0+\text{BSE}@\text{PBE0($\mathrm{\alpha}$)}}
\newcommand{\vv}[1]{\text{#1}_{\vac\vac}}
\def\ti{\vv{Ti}}
\def\mo{\vv{Mo}}
\def\sis{\mathrm{Si_{VV}(S)}}
\def\sit{\mathrm{Si_{VV}(T)}}
\def\gs{\ket{\prescript{3}{0}{A''}}}    % triplet ground state (gs)
\def\es{\ket{\prescript{3}{1}{A''}}}    % triplet excited states (es)
\def\pjt{\ket{\prescript{3}{1}{A}}}     % triplet excited PJT state (pjt) -- this is C1
\def\sin{\ket{\prescript{1}{0}{A'}}}    % singlet state (sin)
\def\mogs{\ket{\prescript{3}{0}{A}}}    % triplet ground state (gs)
\def\moes{\ket{\prescript{3}{1}{A}}}    % triplet excited states (es)
\def\mosin{\ket{\prescript{1}{0}{A}}}    % singlet state (sin)

\def\nv{\text{NV}}
\author{Tyler J. Smart}
\affiliation{Department of Physics, University of California, Santa Cruz, CA, 95064, USA}
\alsoaffiliation{Lawrence Livermore National Laboratory, Livermore, California 94551, USA}
\altaffiliation{Contributed equally to this work}
\author{Kejun Li}
\affiliation{Department of Physics, University of California, Santa Cruz, CA, 95064, USA}
\altaffiliation{Contributed equally to this work}
\author{Junqing Xu}
\affiliation{Department of Chemistry and Biochemistry, University of California, Santa Cruz, CA, 95064, USA}
\author{Yuan Ping}
\affiliation{Department of Chemistry and Biochemistry, University of California, Santa Cruz, CA, 95064, USA}
\email{yuanping@ucsc.edu}

\title[]{
    Intersystem Crossing and Exciton-Defect Coupling of Spin Defects in Hexagonal Boron Nitride
}
\begin{document}

%\begin{abstract}
\section{ABSTRACT}
    Despite the recognition of two-dimensional (2D) systems as emerging and scalable host materials of single photon emitters or spin qubits, uncontrolled and undetermined chemical nature of these quantum defects has been a roadblock to further development.
    Leveraging the design of extrinsic defects can circumvent these persistent issues and provide an ultimate solution.
    Here we established a complete theoretical framework to
    accurately and systematically design quantum defects in wide-bandgap 2D systems. With this approach, essential static and dynamical properties are equally considered for spin qubit discovery.
    In particular, many-body interactions such as defect-exciton couplings are vital for describing excited state properties of defects in ultrathin 2D systems. Meanwhile, nonradiative processes such as phonon-assisted decay and intersystem crossing rates require careful evaluation, which compete together with radiative processes. From a thorough screening of defects based on first-principles calculations, we identify promising single photon emitters such as $\vv{Si}$ and spin qubits such as $\ti$ and $\mo$ in hexagonal boron nitride. This work provided a complete first-principles theoretical framework for defect design in 2D materials.
%\end{abstract}

%%%%%%%%%%%%%%%%%%%%%%%%%%%%%%%%%%%%%%%%%%%%%%%%%%%%%%%%%%%%%%%%%%%%%%%%%%%%%%%%%%%%%%%%%%%%%%%%%%%%%%%%%%%%%%%%%%
\section{INTRODUCTION}
%%%%%%%%%%%%%%
Optically addressable defect-based qubits offer a distinct advantage in their ability to operate with high fidelity under room temperature conditions \cite{koehl2011room,weber2010quantum}.
Despite tremendous progress made in years of research, systems which exist today remain inadequate for real-world applications. The identification of stable single photon emitters in 2D materials has opened up a new playground for novel quantum phenomena and quantum technology applications, with improved scalability in device fabrication and
a leverage in doping spatial control, qubit entanglement, and qubit tuning \cite{liu20192d,aharonovich2017quantum}.
In particular, hexagonal boron nitride ($h$-BN) has demonstrated that it can host stable defect-based single photon emitters (SPEs) \cite{mendelson2020strain,feldman2019phonon,yim2020polarization,mackoit2019carbon} and spin triplet defects \cite{kianinia2020generation,turiansky2020spinning}.
However,
%whereas the hallmark nitrogen-vacancy ($\nv$) center in diamond has been investigated extensively \cite{gali2019ab},
persistent challenges must be resolved before 2D quantum defects can become the most promising quantum information platform. These challenges include the undetermined chemical nature of
existing SPEs\cite{li2017nonmagnetic, yim2020polarization}, difficulties in controlled generation of desired spin defects, and scarcity of reliable theoretical methods which
can accurately predict critical physical parameters for defects in 2D materials due to their complex many-body interactions.

To circumvent these challenges, design of promising spin defects by high-integrity theoretical methods is urgently needed.
Introducing extrinsic defects can be unambiguously produced and controlled, which fundamentally solves the current issues of undetermined chemical nature of existing SPEs in 2D systems.
% As highlighted by Refs. \citenum{tran2016quantum, exarhos2019magnetic, ivady2018first, van2010model}, promising spin qubit candidates should satisfy several essential criteria: deep defect levels, stable high spin states, large zero-field splitting, efficient radiative recombination, high intersystem crossing rates and long spin coherence and relaxation time.
As highlighted by Ref. \citenum{weber2010quantum,ivady2018first}, promising spin qubit candidates should satisfy several essential criteria: deep defect levels, stable high spin states, large zero-field splitting, efficient radiative recombination, high intersystem crossing rates and long spin coherence and relaxation time.
Using these criteria for theoretical screening can effectively identify promising candidates but
%reliable predictions of these properties
requires theoretical development of first-principles methods, significantly beyond the static and mean-field level. For example, accurate defect charge transition levels in 2D materials necessitates careful treatment of defect charge corrections for removal of spurious charge interactions~\cite{Komsa2015,Wang2015,wu2017first} and electron correlations for non-neutral excitation, e.g.\ from GW approximations~\cite{wu2017first,govoni2015large} or Koopmans-compliant hybrid functionals \cite{smart2018,Nguyen2018,Weng2018,miceli2018nonempirical}. Optical excitation and exciton radiative lifetime must account for defect-exciton interactions, e.g.\ by solving the Bethe-Salpeter equation, due to large exciton binding energies in 2D systems \cite{Sivan2018,gao2020radiative}.  Spin-phonon relaxation time calls for a general theoretical approach to treat complex symmetry and state degeneracy of defective systems, along the line of recent development based on
ab-initio density matrix approach \cite{xu2020spin}.
Spin coherence time due to the nuclei spin and electron spin coupling
can be accurately predicted for defects in solids by combining first-principles and spin Hamiltonian approaches \cite{seo2016quantum,ye2019spin}. In the end, nonradiative processes, such as phonon-assisted nonradiative recombination, have been recently computed with first-principles electron-phonon couplings for defects in $h$-BN \cite{wu2019carrier}, and resulted in less competitive rates than corresponding radiative processes. However, the spin-orbit induced intersystem crossing as the key process for pure spin state initialization during qubit operation has not been investigated for spin defects in 2D materials from first-principles in-depth.
%that enables the initialization of a pure spin state for qubit operation, and has not been investigated for spin defects in 2D materials from first-principles.

\begin{figure}[H]
    \centering
    \includegraphics[keepaspectratio=true,width=0.8\linewidth]{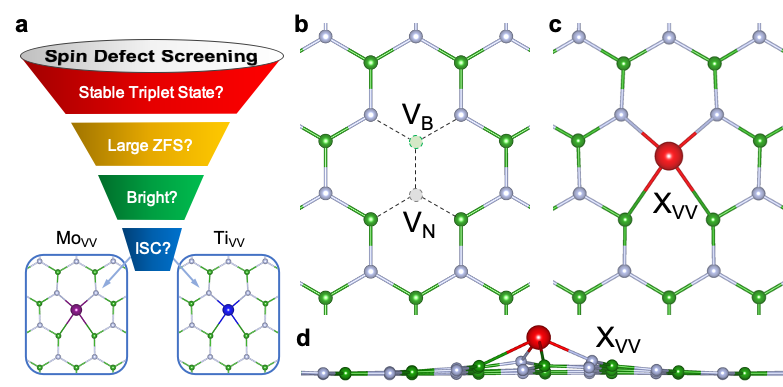}
    \caption{\textbf{Screening of spin defects in $h$-BN.} \textbf{a} Schematic of the screening criteria and workflow developed in this work, where we first search for defects with stable triplet ground state, followed by large zero-field splitting (ZFS), then ``bright'' optical transitions between defect states required for SPEs or qubit operation by photon, and at the end large intersystem crossing rate (ISC) critical for pure spin state initialization. \textbf{b} Divacancy site in $h$-BN corresponding to adjacent B and N vacancies (denoted by $\vb$ and $\vn$). \textbf{c} Top-view and \textbf{d} Side-view of a typical doping configuration when placed at the divacancy site, denoted by $\vv{X}$. Atoms are distinguished by color: grey=N, green=B, purple=Mo, blue=Ti, red=X (a generic dopant).}
    \label{fig:screen}
\end{figure}

This work has developed a complete theoretical framework which enables the design of spin defects based on the critical physical parameters mentioned above and highlighted in Figure~\ref{fig:screen}a. We employed state-of-the-art first-principles methods, focusing on many-body interaction such as defect-exciton couplings and dynamical processes through radiative and nonradiative recombinations. We developed methodology to compute nonradiative intersystem crossing rates with explicit overlap of phonon wavefunctions beyond current implementations in the Huang-Rhys approximation\cite{thiering2017ab}. We showcase the discovery of transition metal complexes such as Ti and Mo with vacancy ($\ti$ and $\mo$) to be spin triplet defects in $h$-BN, and the discovery of $\vv{Si}$ to be a bright SPE in $h$-BN.
% \textcolor{blue}{$\ti$ and $\mo$ are extrinsic defects in $h$-BN} predicted as stable triplets with large zero-field splitting and spin-selective decay, which will set 2D quantum defects at a competitive stage with $\nv$ center in diamond for quantum technology applications.
We predict $\ti$ and $\mo$ are stable triplet defects in $h$-BN (which is rare considering the only known such defect is $\vb^-$~\cite{gottscholl2020initialization}) with large zero-field splitting and spin-selective decay, which will set 2D quantum defects at a competitive stage with $\nv$ center in diamond for quantum technology applications.

%%%%%%%%%%%%%%%%%%%%%%%%
% Figure for screening

%%%%%%%%%%%%%%%%%%%%%%%%%%%%%%%%%%%%%%%%%%%%%%%%%%%%%%%%%%%%%%%%%%%%%%%%%%%%%%%%%%%%%%%%%%%%%%%%%%%%%%%%%%%%%%%%%%
% \section{RESULTS AND DISCUSSION}
\section{RESULTS}

In the development of spin qubits in 3D systems (e.g.\ diamond, SiC, and AlN), defects beyond $sp$ dangling bonds from N or C have been explored. In particular, large metal ions plus anion vacancy in AlN and SiC were found to have potential as qubits due to triplet ground states and large zero-field splitting (ZFS) \cite{seo2017designing}. %For example, Seo et al.\ shows Zr, Hf-N vacancy pairs in AlN show promising properties as qubits
Similar defects may be explored in 2D materials \cite{turiansky2019dangling}, such as the systems shown in Figure~\ref{fig:screen}b-d.
This opens up the possibility of overcoming the current limitations of uncontrolled and undetermined chemical nature of 2D defects, and unsatisfactory spin dependent properties of existing defects.
In the following, we will start the computational screening of spin defects with static properties of the ground state (spin state, defect formation energy and ZFS) and the excited state (optical spectra), then we will discuss dynamical properties including radiative and nonradiative (phonon-assisted spin conserving and spin flip) processes, as the flow chart shown in Figure~\ref{fig:screen}a. We will summarize the complete defect discovery procedure and discuss the outlook at the end.
\subsection{Screening Triplet Spin Defects in $h$-BN}
To identify stable qubits in $h$-BN, we start from screening neutral dopant-vacancy defects for a triplet ground state based on total energy calculations of different spin states at both semi-local PBE (Perdew–Burke-Ernzerhof) and hybrid functional levels. We considered the dopant substitution at a divacancy site in $h$-BN (Figure~\ref{fig:screen}b) for four different elemental groups. The results of this procedure are summarized in Supplementary Table 1 and Note 1. With additional supercell tests in Supplementary Table 2, our screening process finally yielded that only $\mo$ and $\ti$ have a stable triplet ground state.
%\textbf{Thermodynamic Charge Formation Energy}
We further confirmed the thermodynamic charge stability of these defect candidates via calculations of defect formation energy and charge transition levels.
% As shown in SI Figure S1,
As shown in Supplementary Figure 1,
both $\ti$ and $\mo$ defects have a stable neutral ($q=0$) region for a large range of Fermi level ($\varepsilon_F$), from 2.2 eV to 5.6 eV for $\mo$ and from 2.9 eV to 6.1 eV for $\ti$. These neutral states will be stable in intrinsic $h$-BN systems or with weak p-type or n-type doping (see Supplementary Note 2).

With a confirmed triplet ground state, we next computed the two defects' zero-field splitting. A large ZFS is necessary to isolate the $m_s = \pm 1$ and $m_s = 0$  levels even at zero magnetic field allowing for controllable preparation of the spin qubit.
Here we computed the contribution of spin-spin interaction to ZFS by implementing the plane-wave based method developed by Rayson et al.\ (see Methods section for details of implementation and benchmark on $\nv$ center in diamond) \cite{rayson2008first}. Meanwhile, the spin-orbit contribution to ZFS was computed with the ORCA code.
We find that both defects have sizable ZFS including both spin-spin and spin-orbit contributions (axial $D$ parameter) of 19.4 GHz for $\ti$ and 5.5 GHz for $\mo$, highlighting the potential for the basis of a spin qubit with optically detected magnetic resonance (ODMR) (see Supplementary Note 3 and Figure 2).
They are notably larger than previously reported values for ZFS of other known spin defect in solids \cite{seo2017designing}, although at a reasonable range considering large ZFS values (up to 1000 GHz) in transition-metal complex molecules \cite{zolnhofer2020electronic}.

\subsection{Screening SPE Defects in $h$-BN}
To identify single photon emitters in $h$-BN, we considered a separate screening process of these dopant-vacancy defects, targeting those with desirable optical properties. Namely, an SPE efficiently emits a single photon at a time at room temperature.
%corresponding to narrow zero-phonon emission.
Physically this corresponds to identifying defects which have a single bright intra-defect transition with high quantum efficiency (i.e.\ much faster radiative rates than nonradiative ones), for example current SPEs in $h$-BN have radiative lifetimes $\sim$1-10 ns and quantum efficiency over 50$\%$.
\cite{tran2016robust,schell2017coupling}

Using these criteria we screened the defects by computing their optical transitions and radiative lifetime at Random Phase Approximation (RPA) (see Supplementary Note 4, Figure 3 and Table 3). This offers a cost-efficient first-pass to identify defects with bright transition and short radiative lifetime as potential candidates for SPEs.
%As the defects that can act as bright SPEs are experimentally observable and worthy of being studied, we additionally performed Random Phase Approximation (RPA) with PBE state, which is similar to experimental absorption spectra, to screen SPE defects. The first peaks that are related to defect-defect transition in the RPA spectra are under consideration.
From this procedure, we found that $\vv{C}$(T), $\sis$, $\sit$, $\vv{S}$(S), $\vv{Ge}$(S) and $\vv{Sn}$(S) could be promising SPE defects ((T) denotes triplet; (S) denotes singlet), with a bright intra-defect transition and radiative lifetimes on the order of 10 ns, at the same order of magnitude of the SPEs' lifetime observed experimentally.
\cite{schell2017coupling}
Among these, $\sis$ has the shortest radiative lifetime, and in addition, Si has recently been experimentally detected in $h$-BN with samples grown in chemical vapor deposition (the ground state of $\vv{Si}$ is also singlet).\cite{ahmadpour2019substitutional}
% $\vv{Ge}$(S) and $\vv{Sn}$(S) are not as bright as $\sis$; $\vv{C}$(T) has already been the subject of a recent study \cite{ali2020vncb} which compares its photoluminescence properties to experimental SPEs; meanwhile we found that $\vv{S}$(S) has an excitation energy too far in the UV region.
Hence we will focus on $\vv{Si}$ as an SPE candidate in the following sections as we compute optical and electronic properties at higher level of theory from many-body perturbation theory including accurate electron correlation and electron-hole interactions.
%We will recompute the optical and electronic properties from many-body perturbation theory including accurate electron correlation and electron-hole interactions for both spin qubit ($\ti$, $\mo$) and SPE ($\vv{Si}$) candidates.
Note that $\vv{C}$ (commonly denoted $\rm C_BV_N$) has also been suggested to be a SPE source in $h$-BN.~\cite{sajid2020vncb}

% \textcolor{tcolor}{\textbf{TS: Why did you choose to look at Si but not S or C?}}
% \textcolor{tcolor}{\textbf{TS: This part still needs to be better finished. \textcolor{blue}{K: done, please take a look and welcome any comments}}}

\subsection{Single-Particle Levels, Optical Spectra and Radiative Lifetime}

%Deep defect levels with an allowed isolated and spin-polarized optical transition is another key requirement for potential defects as single photon emitters and spin qubits. Therefore, we will next investigate the electronic structure and optical spectra of the \textcolor{tcolor}{aforementioned defects from many-body perturbation theory including excitonic effects.}

\begin{figure}[H]
    \centering
    \includegraphics[keepaspectratio=true,width=1.0\linewidth]{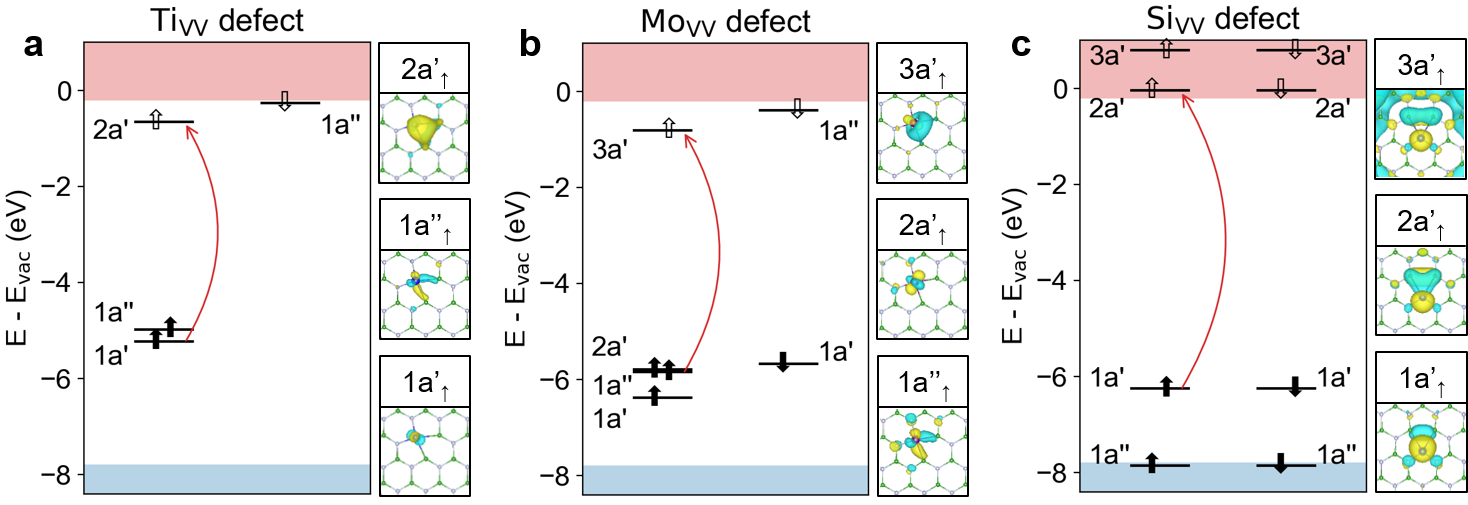}
    \caption{
        \textbf{Single-particle levels and wavefunctions.}
    Single-particle defect levels (horizontal black lines) of the (a) $\ti$, (b) $\mo$, and (c) $\vv{Si}$ defects in $h$-BN, calculated at $\mathrm{G_0W_0}$ with PBE0($\alpha$) starting wavefunctions. The blue/red area corresponds to the valence/conduction band of $h$-BN.
    States are labelled by their ordering and representation within the $C_S$ group with up/down arrows indicating spin and filled/unfilled arrows indicating occupation.
    A red arrow is drawn to denote the intra-defect optical transition found in Figure~\ref{fig:bse}.
    Defect wavefunctions at PBE0($\alpha$) are shown with an isosurface value 10\% of the maximum. The blue and yellow color denotes different signs of wavefunctions.
    }
    \label{fig:sg_ctl}
\end{figure}

The single-particle energy levels of $\ti$, $\mo$ and $\vv{Si}$ are shown in Figure~\ref{fig:sg_ctl}.
These levels are computed by many-body perturbation theory ($\mathrm{G_0W_0}$) for accurate electron correlation,
with hybrid functional (PBE0($\alpha$), $\alpha=0.41$ based on the Koopmans' condition~\cite{smart2018}) as the starting point to address self-interaction errors for $3d$ transition metal defects.
\cite{fuchs2007quasiparticle,bechstedt2016many}.
For example, we find that both the wavefunction distribution and ordering of defect states can differ between PBE and PBE0($\alpha$) (see Supplementary Figure 4-6). The convergence test of $\mathrm{G_0W_0}$ can been found in Supplementary Figure 7, Note 5, and Table 4.
Importantly, the single particle levels in Figure~\ref{fig:sg_ctl} show there are well localized occupied and unoccupied defect states in the $h$-BN band gap, which yield the potential for intra-defect transitions.

Obtaining reliable optical properties of these two-dimensional materials necessitates solving the Bethe-Salpeter equation (BSE)
%, \textcolor{red}{make a note here: see numerical convergence in SI Figure S8-S10}
to include excitonic effects due to their strong defect-exciton coupling, which is not included in RPA calculations (see comparison in Supplementary Figure 8 and Table 5). \cite{ping2013,PRBDario-2012,Ping2012-prb,Ping-2013WO3}
The BSE optical spectra are shown for each defect in Figure~\ref{fig:bse}a-c (the related convergence tests can be found in Supplementary Figure 9-10).
In each case we find an allowed intra-defect optical transition (corresponding to the lowest energy peak as labeled in Figure~\ref{fig:bse}a-c, and red arrows in Figure~\ref{fig:sg_ctl}).
From the optical spectra we can compute their radiative lifetimes as detailed in the Methods section on Radiative Recombination.
We find the transition metal defects' radiative lifetimes (tabulated in Table~\ref{table:rad}) are long, exceeding $\mu$s.
% Therefore, they are not good candidates for SPEs, although they still have potentials for spin qubits with optically-allowed intra-defect transitions.
Therefore, they are not good candidates for SPE. In addition, while they still are potential spin qubits with optically-allowed intra-defect transitions, optical readout of these defects will be difficult.
Referring to Table~\ref{table:rad} and the expression of radiative lifetime in Eq.~\ref{eq:rate-radiative-0D} we can see this is due to their low excitation energies ($E_0$, in the infrared region) and small dipole moment strength ($\mu^2_{e-h}$). The latter is related to the tight localization of the excitonic wavefunction for $\ti$ and $\mo$ (shown in Figure~\ref{fig:bse}d-f), as strong localization of the defect-bound exciton leads to weaker oscillator strength~\cite{hours2005exciton}.

\begin{figure}[ht!]
    \centering
    \includegraphics[keepaspectratio=true,width=1.0\linewidth]{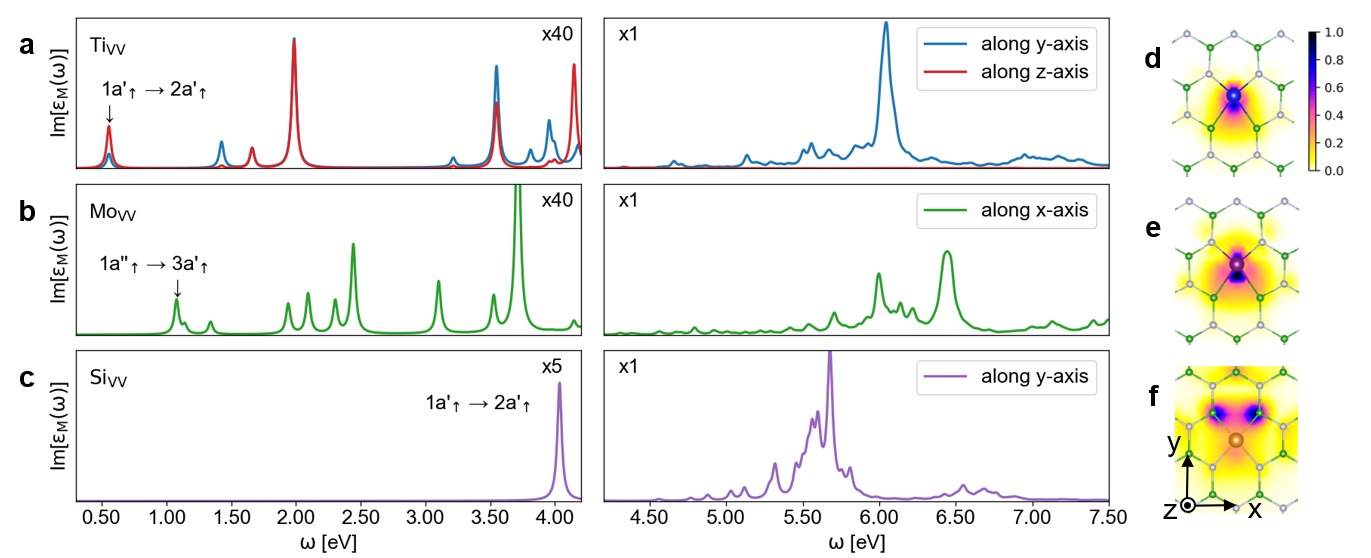}
    \caption{
        \textbf{BSE optical spectra and exciton wavefunctions.}
    Absorption spectra  of the (a) $\ti$, (b) $\mo$, and (c) $\vv{Si}$ defects in $h$-BN at the level of $\gwbse$. The left and right panels provide absorption spectra for two different energy ranges, where the former is magnified by a factor of 40 for $\ti$ and $\mo$ and a factor of 5 for $\vv{Si}$ for increasing visibility. A spectral broadening of 0.02 eV is applied.
    %The absorption spectra of pristine $h$-BN is computed with light polarized in the plane.
    %The \textcolor{blue}{pictures in the first column} show the exciton wavefunctions for the first peak.
    The exciton wavefunctions of (d) $\ti$, (e) $\mo$ and (f) $\vv{Si}$ are shown on the right for the first peak.
    }
    \label{fig:bse}
\end{figure}

On the other hand, the optical properties of the $\vv{Si}$ defect are quite promising for SPEs, as Figure~\ref{fig:bse}c shows it has a very bright optical transition in the ultraviolet region. As a consequence, we find that the radiative lifetime (Table~\ref{table:rad}) for $\vv{Si}$ is 22.8 ns at $\rm G_0W_0+BSE@PBE0(\alpha)$.
We note that although the lifetime of $\vv{Si}$ at the level of BSE is similar to that obtained at RPA (13.7 ns),
the optical properties of 2D defects at RPA are still unreliable, due to the lack of excitonic effects. For example, the excitation energy ($E_0$) can deviate by $\sim$1 eV and oscillator strengths ($\mu_{e-h}^2$) can deviate by an order of magnitude (more details can be found in Supplementary Table 5).
Above all, the radiative lifetime of $\vv{Si}$ is comparable to experimentally observed SPE defects in $h$-BN,\cite{schell2017coupling} showing that $\vv{Si}$ is a strong SPE defect candidate in $h$-BN.

\begin{table}[H]
    \footnotesize
    \centering

\begin{tabular}{c c c c c}
    \hline \hline
        Defect    &   $E_{0}$ (eV)    &   $\mu^2_{e-h}$ (bohr$^2$)   &    $\tau_R$ (ns)   &   $E_{b}$ (eV)\\
    \hline
         $\ti$    &       0.556       &        $2.81*10^{-2}$        &    $1.95*10^{5}$   &   4.018\\
         $\mo$    &       1.079       &        $2.29*10^{-2}$        &    $3.26*10^{4}$   &   3.965\\
         $\vv{Si}$   &       4.036       &        $6.28*10^{-1}$        &    $22.8$          &   2.189\\
         %$\sit$   &       2.883       &        $5.83*10^{-1}$        &    $67.3$          &   3.099\\
        $\nbvn$   &       2.408       &        $1.87$                &    $35.9$          &   2.428\\
    \hline \hline
\end{tabular}

    \caption{Optical excitation energy ($E_{0}$), modulus square of the transition dipole moment ($\mu^2_{e-h}$), radiative lifetime ($\tau_R$) and exciton binding energy ($E_b$) of several defects in $h$-BN at the level of theory of $\gwbse$. The corresponding excitation transitions are $1a'_\uparrow \rightarrow 2a'_\uparrow$ for the $\ti$ defect, $1a''_\uparrow \rightarrow 3a'_\uparrow$ for the $\mo$ defect and $1a'_\uparrow \rightarrow 2a'_\uparrow$ for the $\vv{Si}$ defect. For comparison, we include the results of $\nbvn$ (in-plane structure) from Ref.~\citenum{wu2019carrier}.
    }
    \label{table:rad}
\end{table}
%%%%%%%%%%

%%%%%%%%%%%%%%%%%%%%%%%%%%%%%%%%%%%%%%%%%%%%%%%%%%%%%%%%%%%%%%%%%%%%%%%%%%%%%%%%%%%%%%%%%%%%%%%%%%%%%%%%%%%%%%%%%%

%\subsection{Multiplet Structure and Dynamics of the Ti Defect}

% \begin{itemize}
%     \item Radiative pathways (Table: radiative, nonradiative, and nonradiative has different processes. Indexed by transition)
%     \item Nonradiative pathway
%         \subitem Pseudo Jahn-Teller effects in the excited singlet and triplet states (discuss its participation in nonradiative recombination)
%         \subitem Intersystem crossing/SOC coupling constants
%     \item Final notes, looking at the whole picture, what makes Ti a good/novel spin qubit. Static properties are good (band structure and ZFS) but also dynamical properties (PJT and ISC)
% \end{itemize}

\subsection{Multiplet Structure and Excited-State Dynamics}
% Finally, we discuss the excited-state dynamics of $\ti$ defect in $h$-BN, which can allow for polarization of the system to a particular spin state by optical pumping and develop strategies for realistic spin qubit operation.
Finally, we discuss the excited-state dynamics of the spin qubit candidates $\ti$ and $\mo$ defects in $h$-BN, where the possibility of intersystem crossing is crucial.
This can allow for polarization of the system to a particular spin state by optical pumping, required for realistic spin qubit operation.

\begin{figure}[H]
    \centering
    \includegraphics[keepaspectratio=true,width=1\linewidth]{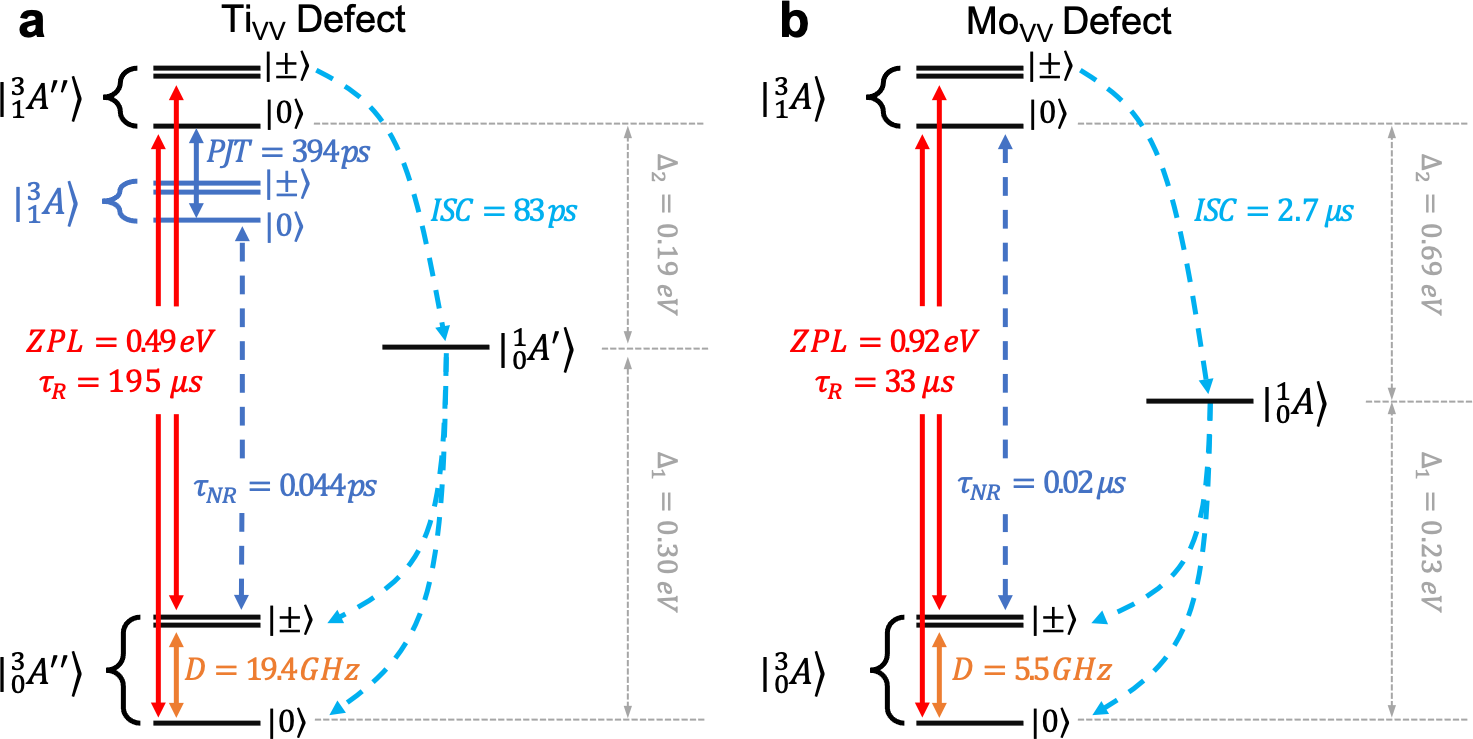}
    \caption{
        \textbf{Multiplet structure of triplet defects.}
        Multiplet structure and related radiative and nonradiative recombination rates of the (\textbf{a}) $\ti$ defect and the (\textbf{b}) $\mo$ defect in $h$-BN, computed at $T=10K$.
    The radiative process is shown in red with zero-phonon line (ZPL) and radiative lifetime ($\tau_R$); the ground state nonradiative recombination ($\tau_{NR}$) is denoted with a dashed line in dark blue; and finally the intersystem crossing (ISC) to the singlet state from the triplet excited state is shown in light blue. The zero-field splitting ($D$) is denoted by the orange line. For the $\ti$ defect, the pseudo Jahn-Teller (PJT) process is shown with a solid line in dark blue.
    }
    \label{fig:multi}
\end{figure}

An overview of the multiplet structure and excited-state dynamics is given in Figure~\ref{fig:multi} for the $\ti$ and $\mo$ defects.
For both defects, the system will begin from a spin-conserved optical excitation from the triplet ground state to the triplet excited state, where next the excited state relaxation and recombination can go through several pathways. The excited state can directly return to the ground state via a radiative (red lines) or nonradiative process (dashed dark blue lines). For the $\ti$ defect shown in Figure~\ref{fig:multi}a, we find the system may relax to another excited state with lower symmetry through a pseudo Jahn-Teller distortion (PJT; solid dark blue lines), and ultimately recombine back to the ground state nonradiatively. Most importantly, a third pathway is to nonradiatively relax to an intermediate singlet state through a spin-flip intersystem crossing (ISC), and then again recombine back to the ground state (dashed light-blue lines).
% \textcolor{blue}{The $\mo$ defect possesses similar processes except for the PJT process due to its lack of symmetry ($C_1$).}
This ISC pathway is critical for the preparation of a pure spin state, similar to $\nv$ center in diamond.
Below, we will discuss our results for the lifetime of each radiative or nonradiative process, in order to determine the most competitive pathway under the operation condition.

\subsection{Direct Radiative and Nonradiative Recombination}
%($\es \rightarrow \gs$)} %Extending from our discussion on the electronic structure and optical properties of spin defects in $h$-BN,

\begin{figure}[H]
    \centering
    \includegraphics[keepaspectratio=true,width=0.8\linewidth]{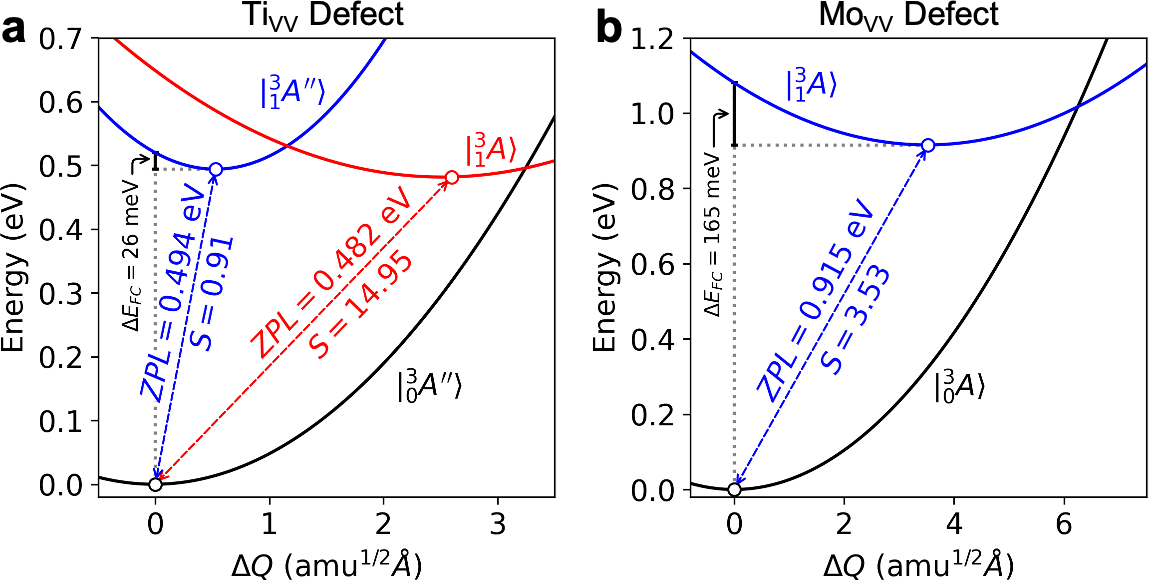}
    \caption{
        \textbf{Configuration coordinate diagrams.}
    Configuration diagram of the (\textbf{a}) $\ti$ defect and (\textbf{b}) $\mo$ defect in $h$-BN. The potential energy surfaces for each state are as follows: the triplet ground state in black, triplet excited state in blue, and for the $\ti$ defect the pseudo Jahn-Teller triplet excited state in red. The zero-phonon lines (ZPL) are given as the energetic separation between the minima of the respective potential energy surfaces, along with the corresponding Huang-Rhys factors ($S$). The dashed black line represents the vertical excitation energy between triplet ground and excited states, and $\Delta E_{FC}$ represents relaxation energy to equilibrium geometry at the excited state.
    }
    \label{fig:config}
\end{figure}

% \textcolor{red}{this section may need to be better organized}
First, we will consider the direct ground state recombination processes.
Figure~\ref{fig:config} shows the configuration diagram of the $\ti$ and $\mo$ defects.
% ground state triplet configuration $\gs$ of $\ti$ alongside the first triplet excited state $\es$. Likewise, the $\ti$ defect is shown with ground state $\mogs$ and excited state $\moes$.
The zero-phonon line (ZPL) for direct recombination can be accurately computed by subtracting its vertical excitation energy computed at BSE (0.56 eV for $\ti$ and 1.08 eV for $\mo$) by its relaxation energy in the excited state (i.e.\ Franck-Condon shift~\cite{van2004first}, $\Delta E_{FC}$ in Figure~\ref{fig:config}). This yields ZPLs of 0.53 eV and 0.91 eV for $\ti$ and $\mo$, respectively. Although this method accurately includes both many-body effects and Franck-Condon shifts, it is difficult to evaluate ZPLs for the triplet to singlet-state transition currently.
%computing ZPLs by this method is currently limited (for example, we cannot evaluate energy separation of triplet/singlet states in this way).
Therefore, we compared with the ZPLs computed by constrained occupation DFT (CDFT) method at PBE. This yields ZPLs of 0.49 eV and 0.92 eV for $\ti$ and $\mo$, respectively, which are in great agreement with the ones obtained from BSE excitation energies subtracting $\Delta E_{FC}$ above.
%\textcolor{blue}{Considering the system is neutral in CDFT, electron and hole pairs are included implicitly, and the good comparison to BSE may imply it is not a bad approximation for exciton-phonon coupling. Recently, there are promising developments of explicit exciton-phonon coupling via BSE which can compute exciton recombination by phonons~\cite{chen2020exciton}; however, further development is desired to include multi-phonon process or compute total energy and force at BSE for effective phonon estimation needed for these defect processes.}
Lastly, the radiative lifetimes for these transitions are presented in Table~\ref{table:rad} as discussed in the earlier section, which shows $\ti$ and $\mo$ have radiative lifetimes of 195 $\mu$s and 33 $\mu$s, respectively (red lines in Figure~\ref{fig:multi}).

In terms of nonradiative properties, the small Huang-Rhys ($S_f$) for the $\es$ to $\gs$ transition of the $\ti$ defect (0.91) implies extremely small electron-phonon coupling and potentially an even slower nonradiative process.
On the other hand, $S_f$ for the $\moes$ to $\mogs$ transition of the $\mo$ defect is modest (3.53) and may indicate a possible nonradiative decay.
%%%%%%%%%%%%%%%%%%%%%%
%%%%%%%%%%%%%%%%%%%%%%
Following the formalism presented in Ref.~\citenum{wu2019carrier}, we computed the nonradiative lifetime of the ground state direct recombination ($T=10$ K is chosen to compare with measurement at cryogenic temperatures \cite{goldman2015phonon}). %for the $\es$ triplet excited state to the $\gs$ triplet ground state.
Consistent with their Huang-Rhys factors, the nonradiative lifetime of $\ti$ is found to be 10 s, while the nonradiative lifetime of the $\mo$ defect is found to be 0.02 $\mu$s.
The former lifetime is indicative of a forbidden transition; however, the $\ti$ defect also possesses a pseudo Jahn-Teller (PJT) effect in the triplet excited state (red curve in Figure~\ref{fig:config}a).
Due to the PJT effect, the excited state ($C_S$, $\es$) can relax to lower symmetry ($C_1$, $\pjt$) with a nonradiative lifetime of 394 ps (solid dark blue line in Figure~\ref{fig:multi}a, additional details see Supplementary Note 9 and Figure 11). Afterward, nonradiative decay from $\pjt$ to the ground state ($\gs$) (dashed dark blue line in Figure~\ref{fig:multi}a) exhibits a lifetime of 0.044 ps due to a large Huang-Rhys factor (14.95).

%%%%%%%%%%%%%%%%%%%%%%%
\subsection{Spin-Orbit Coupling and Nonradiative Intersystem Crossing Rate}
%($\es \rightarrow \sin$)}
Lastly, we considered the possibility of an ISC between the triplet excited state and the singlet ground state for each defect, which is critical for spin qubit application.
In order for a triplet to singlet transition to occur, a spin-flip process must take place. For ISC, typically spin-orbit coupling (SOC) can entangle triplet and singlet states yielding the possibility for a spin-flip transition. To validate our methods for computing SOC (see methods section), we first computed the SOC strengths for the $\nv$ center in diamond. We obtained SOC values of 4.0 GHz for  the axial $\lambda_z$ and 45 GHz for non-axial  $\lambda_\perp$ in fair agreement with previously computed values and experimentally measured values \cite{thiering2017ab,bassett2014ultrafast}.
We then computed the SOC strength for the $\ti$ defect ($\lambda_z = 149$ GHz, $\lambda_\perp = 312$ GHz) and the $\mo$ defect ($\lambda_z = 16$ GHz, $\lambda_\perp = 257$ GHz).
The value of $\lambda_\perp$ in particular leads to the potential for a spin-selective pathway for both defects, analogous to $\nv$ center in diamond.

% To compute the ISC rate, we considered a new approach which is a derivative of the nonradiative recombination formalism presented in Eq.~\ref{eq:rate-nonrad-allterms}:
To compute the ISC rate, we developed an approach which is a derivative of the nonradiative recombination formalism presented in Eq.~\ref{eq:rate-nonrad-allterms}:
\begin{align}
    \Gamma_{ISC} &= 4 \pi \hbar \lambda_\perp^2 \widetilde{X}_{if}(T) \label{eq:isc-full}\\
    \widetilde{X}_{if}(T) &= \sum_{n,m}p_{in} \left|
        \braket{\phi_{fm}(\textbf{R})}{\phi_{in}(\mathbf{R})}
    \right|^2 \delta(m\hbar\omega_f - n\hbar\omega_i+\Delta E_{if}) \label{eq:isc-F}
\end{align}
% Opposed to Eq.~\ref{eq:rate-nonrad-X}, the phonon overlap in Eq.~\ref{eq:isc-F} does not have an expectation on $\Delta Q$, however it behaves similarly nonetheless and has an exponential dependence on the Huang-Rhys factor $S$.
Compared with previous formalism, \cite{thiering2017ab} this method
%has the advantages of a temperature dependence (previous approximate $T \rightarrow 0$) and
allows different values for initial state vibrational frequency ($\omega_i$) and final state one ($\omega_f$) through explicit calculations of phonon wavefunction overlap. Again to validate our methods we first computed the intersystem crossing rate for $\nv$ center in diamond. Using the experimental value for $\lambda_\perp$ we obtain an intersystem crossing rate for $\nv$ center in diamond of 2.3 MHz which is in excellent agreement with the experimental value of 8 and 16 MHz \cite{goldman2015phonon}.
In final, we obtain an intersystem crossing time of 83 ps for $\ti$ and 2.7 $\mu$s for $\mo$ as shown in Table~\ref{table:nonrad} and light blue lines in Figure~\ref{fig:multi}.

% Table for nonradiative lifetimes
\begin{table}
    \footnotesize
    \centering

% \resizebox{1\columnwidth}{!}{
\begin{tabular}{cccccc}
\hline \hline
    % \multirow{2}{*}{GSR}
    $\ti$ & GSR & ZPL (eV) & $S_f$ & $W_{if}$ (eV/(amu$^{1/2}\si{\angstrom}$)) & $\tau_{NR}$ (ps) \\
    % & & (eV) & & eV/(amu$^{1/2}\si{\angstrom}$) & (ps) \\

    & $\es \rightarrow \gs$  & 0.494 & 0.91 & $1.02\times 10^{-1}$ & $8.80\times 10^{12}$ \\
    & $\pjt \rightarrow \gs$ & 0.482 & 14.95 & $1.91\times 10^{-2}$ & $4.41\times 10^{-2}$ \\

    & PJT & $E_{JT}$ (eV) & $S_f$  & $\delta_{JT}$ (eV) & $\tau^C_{NR}$ (ps) \\
    % &  & (eV) & & (eV) & (ps)\\

    % $\es \rightarrow \pjt$ & 0.012 & 13.30 & -- & 10.75 & 0.006 & -- & $3.94\times 10^{2}$ \\
    & $\es \rightarrow \pjt$ & 0.012 & 10.75 & 0.006 & $3.94\times 10^{2}$ \\

    & ISC & ZPL (eV) & $S_f$ & $\lambda_{\perp}$ (GHz) & $\tau_{ISC}$ (ps) \\
    % & & (eV) & & (GHz) & (ps) \\

    & $\es \rightarrow \sin$ & 0.189 & 17.48 & 312 & $8.30\times 10^{1}$ \\

\hline
    $\mo$ & GSR & ZPL (eV) & $S_f$ & $W_{if}$ (eV/(amu$^{1/2}\si{\angstrom}$)) & $\tau_{NR}$ ($\rm \mu s$) \\
    % & & (eV) & & eV/(amu$^{1/2}\si{\angstrom}$) & ($\rm \mu s$)\\

    & $\moes \rightarrow \mogs$  & 0.915 & 22.05 & $1.5\times 10^{-2}$ & $0.02$ \\

    & ISC & ZPL (eV) & $S_f$ & $\lambda_{\perp}$ (GHz) & $\tau_{ISC}$ ($\rm \mu s$) \\
    % & & (eV) & & (GHz) & ($\rm \mu s$) \\

    & $\moes \rightarrow \mosin$ & 0.682 & 7.22 & 257 & $2.7$ \\

\hline \hline
\end{tabular}
% }

    \caption{Various nonradiative recombination lifetimes along with relevant quantities for the $\ti$ and $\mo$ defects in $h$-BN, including ground state recombination (GSR), pseudo Jahn-Teller (PJT), and intersystem crossing (ISC).
    % \textcolor{tcolor}{The units of $W_{if}$ is eV/(amu$^{1/2}\si{\angstrom}$).}
    %$\Delta Q$ has units of amu$^{1/2}${\AA}, $\hbar\omega$ has units of meV, $W_{if}$ has units of eV$/$(amu$^{1/2}${\AA}), $X_{if}$ has units of (amu\ {\AA}$^2$)/eV, and $\widetilde{X}_{if}$ has units of $1/$eV. Computed at $T=10K$. \textcolor{blue}{where are units for lifetime??}
    }
    \label{table:nonrad}
\end{table}

The results of all the nonradiative  pathways for two spin defects are summarized in Table~\ref{table:nonrad} and are displayed in Figure~\ref{fig:multi} along with the radiative pathway.
We begin by summarizing the results for $\ti$ first and then discuss $\mo$ below. In short, for $\ti$ the spin conserved optical excitation from the triplet ground state $\gs$ to the triplet excited state $\es$ cannot directly recombine nonradiatively due to a weak electron-phonon coupling between these states.
In contrast, a nonradiative decay is possible via its PJT state ($\pjt$) with a lifetime of 394 ps.
% \textcolor{red}{less emphasize PJT passway in next two sen.} In contrast, nonradiative recombination between the PJT state ($\pjt$) and triplet ground state is extremely fast, due to a strong electron-phonon coupling. Yet, the recombination between the PJT state and the triplet ground state is limited by the transition rate from the triplet excited state ($\es$) to the PJT state ($\pjt$) (394 ps).
Finally, the process of intersystem crossing from the triplet excited state $\es$ to the singlet state ($\sin$) is an order of magnitude faster (i.e.\ 83 ps) and is in-turn a dominant relaxation pathway. Therefore the $\ti$ defect in $h$-BN is predicted to have an expedient spin purification process due to a fast intersystem crossing with a rate of 12 GHz. % The interconnected nature of these dynamic processes highlights the importance of a full computational approach as we have presented here.
We note that while the defect has a low optical quantum yield and is predicted to not be a good SPE candidate, it is still noteworthy, as to date the only discovered triplet defect in $h$-BN is the negatively charged boron vacancy, which also does not exhibit SPE and has similarly low quantum efficiency. \cite{kianinia2020generation} Meanwhile, the leveraged control of an extrinsic dopant can offer advantages in spatial and chemical nature of defects.
% \textcolor{red}{I would not call "dark defect"; it meant forbidden transition! call defect with low quantum efficiency}

For the $\mo$ defect, its direct nonradiative recombination lifetime from the triplet excited state $\moes$ to the ground state $\mogs$ is 0.02 $\mu$s.
While the comparison with its radiative lifetime (33 $\mu$s) is improved compared to the $\ti$ defect, it still is predicted to have low quantum efficiency.
However, again the intersystem crossing between $\moes$ and $\mosin$ is competitive with a lifetime of 2.7 $\mu$s. This rate (around MHz) is similar to diamond and implies a feasible intersystem crossing. Owing to its more ideal ZPL position ($\sim$1eV) and improved quantum efficiency, optical control of the $\mo$ defect is seen as more likely and may be further improved by other methods such as coupling to optical cavities~\cite{kim2018photonic,zhong2018optically} and applying strain~\cite{wu2019carrier,mendelson2020strain}.

%%%%%%%%%%%%%%%%%%%%%%%%%%%%%%%%%%%%%%%%%%%%%%%%%%%%%%%%%%%%%%%%%%%%%%%%%%%%%%%%%%%%%%%%%%%%%%%%%%%%%%%%%%%%%%%%%%
% \section{CONCLUSIONS}
\section{DISCUSSION}
In summary, we proposed a general theoretical framework for identifying and designing optically-addressable spin defects for the future development of quantum emitter and quantum qubit systems.
%  , and been the first to investigate transition metal defects in monolayer $h$-BN.
We started from searching for defects with triplet ground state by DFT total energy calculations which allow for rapid identification of possible candidates. Here we found that the $\ti$ and $\mo$ defects in $h$-BN have a neutral triplet ground state. We then
computed zero-field splitting of secondary spin quantum sublevels and found they are sizable for both defects, larger than that of $\nv$ center in diamond, enabling possible control of these levels for qubit operation.
In addition, we screened for potential single photon emitters (SPEs) in $h$-BN based on allowed intra-defect transitions and radiative lifetimes, leading to the discovery of $\vv{Si}$.
Next the electronic structure and optical spectra of each defect were computed from many-body perturbation theory.
% Our results show optically addressable transitions for these defects, with much larger exciton binding energy for the transition metals compared to typical intrinsic $sp$ defects, due to strongly localized exction wavefunctions.
Specifically, the $\vv{Si}$ defect is shown to possess an exciton radiative lifetime similar to experimentally observed SPEs in $h$-BN and is a potential SPE candidate.
Finally, we analyzed all possible radiative and nonradiative dynamical processes with first-principles rate calculations. In particular, we identified a dominant spin-selective decay pathway via intersystem crossing at the $\ti$ defect which gives a key advantage for initial pure spin state preparation and qubit operation.
Meanwhile, for the $\mo$ defect we found that it has the benefit of improved quantum efficiency for more realistic optical control.

%%%%%%%%%%%

This work emphasizes that the theoretical discovery of spin defects requires careful treatment of many-body interactions and various radiative and nonradiative dynamical processes such as intersystem crossing.
We demonstrate high potential of extrinsic spin defects in 2D host materials as qubits for quantum information science.
Future work will involve further examination of spin coherence time and its dominant decoherence mechanism,
as well as other spectroscopic fingerprints from first-principles calculations to facilitate experimental validation of these defects.

%%%%%%%%%%%%%%%%%%%%%%%%%%%%%%%%%%%%%%%%%%%%%%%%%%%%%%%%%%%%%%%%%%%%%%%%%%%%%%%%%%%%%%%%%%%%%%%%%%%%%%%%%%%%%%%%%%
\section{METHODS}
% \begin{itemize}
%     \item Basic info
%     \item Charge formation energy
%     \item Radiative lifetime
%     \item Nonradiative lifetime
% \end{itemize}

\subsection{First-Principles Calculations} In this study, we used the open source plane-wave code Quantum ESPRESSO~\cite{QE} to perform calculations on all structural relaxations and total energies with optimized norm-conserving Vanderbilt (ONCV) pseudopotentials~\cite{ONCV1} and a wavefunction cutoff of 50 Ry. A supercell size of $6\times 6$ or higher was used in our calculations with a $3\times 3\times 1$ k-point mesh. Charged cell total energies were corrected to remove spurious charge interactions by employing the techniques developed in Refs.~\citenum{PingJCP, wu2017first,wang2020layer} and implemented in the JDFTx code \cite{JDFTx}. % \cite{JDFTx, ismail2000new, arias1992ab}.
The total energies, charged defect formation energies and geometry were evaluated at the Perdew-Burke-Ernzerhof (PBE) level~\cite{PBE1997}.
Single-point calculations with k-point meshes of $2\times 2\times 1$ and $3\times 3\times 1$ were performed using hybrid exchange-correlation functional PBE0($\alpha$), where the mixing parameter $\alpha=0.41$ was determined by the generalized Koopmans’ condition as discussed in Ref. \citenum{smart2018, miceli2018nonempirical}. Moreover, we used the YAMBO code \cite{YAMBO} to perform many-body perturbation theory with the GW approximation to compute the quasi-particle correction %~\cite{govoni2015large, ping2013electronic, nguyen2012improving}
using PBE0($\alpha$) eigenvalues and wavefunctions as the starting point.
The random phase approximation (RPA) and Bethe-Salpeter Equation (BSE) calculations were further solved on top of the GW approximation for the electron-hole interaction to investigate the optical properties of the defects, including absorption spectra and radiative lifetime.
%%%%%%%%%

\subsection{Thermodynamic Charge Transition Levels and Defect Formation Energy}
The defect formation energy ($FE_q$) was computed for the $\ti$ and $\mo$ defects following:
\begin{align}
    FE_q(\varepsilon_F) = E_q - E_{pst} + \sum_i \mu_i \Delta N_i + q \varepsilon_F + \Delta_q
    \label{eq:cfe}
\end{align}
where $E_q$ is the total energy of the defect system with charge $q$, $E_{pst}$ is the total energy of the pristine system, $\mu_i$ and $\Delta N_i$ are the chemical potential and change in the number of atomic species $i$, and $\varepsilon_F$ is the Fermi energy. A charged defect correction $\Delta_q$ was computed for charged cell calculations by employing the techniques developed in Ref.~\citenum{wu2017first, PingJCP}. The chemical potential references are computed as $\mu_{Ti} = E_{Ti}^{bulk}$ (total energy of bulk Ti), $\mu_{Mo} = E_{Mo}^{bulk}$ (total energy of bulk Mo),  $\mu_{BN} = E_{BN}^{ML}$ (total energy of monolayer $h$-BN). Meanwhile the corresponding charge transition levels of defects can be obtained from the value of $\varepsilon_F$ where the stable charge state transitions from $q$ to $q'$.
\begin{align}
    \epsilon_{q|q'} = \frac{FE_q - FE_{q'}}{q' - q}
    \label{eq:ctl}
\end{align}

\subsection{Zero-Field Splitting}
The first-order ZFS due to spin-spin interactions was computed for the dipole-dipole interactions of the electron spin:
\begin{align}
    H_{ss} = \frac{\mu_0}{4\pi} \frac{(g_e\hbar)^2}{r^5} \left[
        3(\textbf{s}_1 \cdot \textbf{r})(\textbf{s}_2 \cdot \textbf{r})
        -(\textbf{s}_1 \cdot \textbf{s}_2)r^2
    \right].
\end{align}
Here, $\mu_0$ is the magnetic permeability of vacuum, $g_e$ is the electron gyromagnetic ratio, $\hbar$ is the Planck's constant, $\textbf{s}_1$, $\textbf{s}_2$ is the spin of first and second electron, respectively, and $\textbf{r}$ is the displacement vector between these two electron.
The spatial and spin dependence can be separated by introducing the effective total spin $\textbf{S}=\sum_i \textbf{s}_i$. This yields a Hamiltonian of the form $H_{ss} =  \textbf{S}^T\hat{\textbf{D}}\textbf{S}$, which introduces the traceless zero-field splitting tensor $\hat{\textbf{D}}$. It is common to consider the axial and rhombic ZFS parameters $D$ and $E$ which can be acquired from the $\hat{\textbf{D}}$ tensor:
\begin{align}
    D = \frac{3}{2} D_{zz} \quad \text{and} \quad E = (D_{yy} - D_{xx})/ 2\ .
\end{align}
Following the formalism of Rayson et al.\ \cite{rayson2008first}, the ZFS tensor $\hat{\textbf{D}}$ can be computed with periodic boundary conditions as:
\begin{align}
    D_{ab} = \frac{1}{2}\frac{\mu_0}{4\pi} (g_e\hbar)^2 \sum_{i>j} \chi_{ij}
        \expval{
            \frac{\textbf{r}^2\delta_{ab}-3\textbf{r}_a\textbf{r}_b}{r^5}
        }{
            \Psi_{ij}(\textbf{r}_1, \textbf{r}_2)
        }.
\end{align}
Here the summation on pairs of $i,j$ runs over all occupied spin-up and spin-down states, with $\chi_{ij}$ taking the value $+1$ for parallel spin and $-1$ for anti-parallel spin, and $\Psi_{ij}(\textbf{r}_1,\textbf{r}_2)$ is a two-particle Slater determinant constructed from the Kohn-Sham wavefunctions of the $i$th and $j$th states. This procedure was implemented as a post-processing code interfaced with Quantum ESPRESSO.
To verify our implementation is accurate, we computed the ZFS of the $\nv$ center in diamond which has a well-established result. Using ONCV pseudopotentials, we obtained a ZFS of 3.0 GHz for $\nv$ center, in perfect agreement with previous reported results \cite{seo2017designing}.
For heavy elements such as transition metals, spin-orbit (SO) coupling can have substantial contribution to zero-field splitting. Here, we also computed the SO contribution of the ZFS as implemented in the ORCA code~\cite{neese2012orca,neese2007calculation} (additional details can be found in Supplementary Note 10, Figure 12, and Table 6).

\subsection{Radiative Recombination}
In order to quantitatively study radiative processes, we computed the radiative rate $\Gamma_R$ from Fermi's Golden Rule and considered the excitonic effects by solving BSE \cite{wu2019dimensionality}:
\begin{align}
    \Gamma_R (\textbf{Q}_{ex}) &=
    \frac{2\pi}{\hbar}
    \sum_{q_L, \lambda}
    \left|
        \mel{G,1_{q_L,\lambda}}
        {H^R}
        {S(\textbf{Q}_{ex}),0}
    \right|^2
    \delta(E(\textbf{Q}_{ex}) -\hbar c q_L).
    \label{eq:rate-radiative-full}
\end{align}
% \begin{align}
%     \Gamma_R (\textbf{Q}) &= \frac{4\pi^2e^2}{\hbar c^2V} \Omega(\textbf{Q})^2 \sum_{q, \lambda} \frac{1}{q} \left|
%         \epsilon_{q\lambda} \cdot \mel{G}{\textbf{r}}{S}
%     \right|^2 \delta(\frac{\Omega(\textbf{Q})}{c} - q)
%     \label{eq: rate-radiative-full}
% \end{align}
Here, the radiative recombination rate is computed between the ground state $G$ and the two-particle excited state $S(\textbf{Q}_{ex})$, $1_{q_L,\lambda}$ and 0 denote the presence and absence of a photon, $H^R$ is the electron-photon coupling (electromagnetic) Hamiltonian,  $E(\textbf{Q}_{ex})$ is the exciton energy, and $c$ is the speed of light.
The summation indices in Eq.~\ref{eq:rate-radiative-full} run over all possible wavevector ($q_L$) and polarization ($\lambda$) of the photon.
Following the approach described in Ref.~\citenum{wu2019dimensionality}, the radiative rate (inverse of radiative lifetime $\tau_R$) in SI unit at zero temperature can be computed for isolated defect-defect transitions as:
 %\Gamma_R  &= \frac{4e^2}{3\hbar c^3} \Omega_0^3 |\mu|_{e-h}^2,\\
\begin{equation}
     \Gamma_R = \frac{n_D e^2}{3\pi\epsilon_0\hbar^4 c^3} E_0^3 \mu_{e-h}^2,
    \label{eq:rate-radiative-0D}
\end{equation}
where $e$ is the charge of an electron, $\epsilon_0$ is vacuum permittivity, $E_0$ is the exciton energy at $\textbf{Q}_{ex}=0$, $n_D$ is the reflective index of the host material and $\mu_{e-h}^2$ is the modulus square of exciton dipole moment with length$^2$ unit. Note that Eq.~\ref{eq:rate-radiative-0D} considers defect-defect transitions in the dilute limit; therefore the lifetime formula for zero-dimensional systems embedded in a host material is used~\cite{gupta2018two,mackoit2019carbon} (also considering $n_D$ is unity in isolated 2D systems at the long-wavelength limit). We did not consider the radiative lifetime of $\ti$ defect at a finite temperature because the first and second excitation energy separation is much larger than $kT$. Therefore a thermal average of the first and higher excited states is not necessary and the first excited state radiative lifetime is nearly the same at 10 K as zero temperature.

\subsection{Phonon-Assisted Nonradiative Recombination} In this work, we compute the phonon-assisted nonradiative recombination rate via a Fermi's golden rule approach:
\begin{align}
    \Gamma_{NR} &= \frac{2\pi}{\hbar}g\sum_{n,m}p_{in}\left|\mel{fm}{H^{e-ph}}{in}\right|^2\delta(E_{in}-E_{fm}) \label{eq:rate-nonradiative-fg}
\end{align}
Here, $\Gamma_{NR}$ is the nonradiative recombination rate between electron state $i$ in phonon state $n$ and electron state $f$ in phonon state $m$, $p_{in}$ is the thermal probability distribution of the initial state $\ket{in}$, $H^{e-ph}$ is the electron-phonon coupling Hamiltonian, $g$ is the degeneracy factor and $E_{in}$ is the energy of vibronic state $\ket{in}$. Within the static coupling and one-dimensional (1D) effective phonon approximations, the nonradiative recombination can be reduced to:
\begin{align}
    \Gamma_{NR} &= \frac{2\pi}{\hbar}g|W_{if}|^2 X_{if}(T), \label{eq:rate-nonrad-allterms}\\
    X_{if}(T) &= \sum_{n,m}p_{in}\left|\mel{\phi_{fm}(\textbf{R})}{Q-Q_a}{\phi_{in}(\mathbf{R})}\right|^2 \delta(m\hbar\omega_f - n\hbar\omega_i+\Delta E_{if}), \label{eq:rate-nonrad-X} \\
    W_{if} &= \mel{\psi_i(\mathbf{r}, \textbf{R})}{\frac{\partial H}{\partial Q}}{\psi_f(\mathbf{r},\textbf{R})}\bigg{|}_{\textbf{R}=\textbf{R}_a}. \label{eq:rate-nonrad-W}
\end{align}
Here, the static coupling approximation naturally separates the nonradiative recombination rate into phonon and electronic terms, $X_{if}$ and $W_{if}$, respectively. The 1D phonon approximation introduces a generalized coordinate $Q$, with effective frequency $\omega_i$ and $\omega_f$. The phonon overlap in Eq.~\ref{eq:rate-nonrad-X} can be computed using the quantum harmonic oscillator wavefunctions with $Q-Q_a$ from the configuration diagram (Figure~\ref{fig:config}). Meanwhile the electronic overlap in Eq.~\ref{eq:rate-nonrad-W} is computed by finite difference using the Kohn-Sham orbitals from DFT at the $\Gamma$ point. The nonradiative lifetime $\tau_{NR}$ is given by taking the inverse of rate $\Gamma_{NR}$. Supercell convergence of phonon-assisted nonradiative lifetime is shown in Supplementary Note 11 and Table 7. We validated the 1D effective phonon approximation by comparing the Huang-Rhys factor with the full phonon calculations in Supplementary Table 8.

\subsection{Spin-Orbit Coupling Constant} Spin-orbit coupling (SOC) can entangle triplet and singlet states yielding the possibility for a spin-flip transition. The SOC operator is given to zero-order by \cite{maze2011properties}:
\begin{align}
    H_{so} = \frac{1}{2} \frac{1}{c^2m_e^2} \sum_i \left(
        \nabla_i V \cross \textbf{p}_i \right) \textbf{S}_i \label{eq:soc-full}
\end{align}
where $c$ is the speed of light, $m_e$ is the mass of an electron, $\textbf{p}$ and $\textbf{S}$ are the momentum and spin of electron $i$ and $V$ is the nuclear potential energy. The spin-orbit interaction can be rewritten in terms of the angular momentum $L$ and the SOC strength $\lambda$ as \cite{maze2011properties},
\begin{align}
    H_{so} = \sum_i \lambda_{\perp} (L_{x,i}S_{x,i} + L_{y,i}S_{y,i}) + \lambda_z L_{z,i}S_{z,i}.
\end{align}
where $\lambda_{\perp}$ and $\lambda_z$ denote the non-axial and axial SOC strength, respectively. The SOC strength was computed for the $\ti$ defect in $h$-BN using the ORCA code by TD-DFT \cite{neese2012orca,de2019predicting}.
More computational details can be found in Supplementary Note 10.

\section{DATA AVAILABILITY}
The data that support the findings of this study and the code for the first-principles methods proposed in this study are available from the corresponding author (Yuan Ping) upon reasonable request.

%%TC:ignore
%\begin{acknowledgement}
\section{ACKNOWLEDGEMENT}
    We acknowledge Susumu Takahashi for helpful discussions.
This work is supported by the National Science Foundation under grant no. DMR-1760260, DMR-1956015 and DMR-1747426.
    Part of this work was performed under the auspices of the U.S. Department of Energy by Lawrence Livermore National Laboratory under Contract DE-AC52-07NA27344. TJS acknowledges the LLNL Graduate Research Scholar Program and funding support from LLNL LDRD 20-SI-004.
This research used resources of the Scientific Data and Computing center, a component of the Computational Science Initiative, at Brookhaven National Laboratory under Contract No. DE-SC0012704,
the lux supercomputer at UC Santa Cruz, funded by NSF MRI grant AST 1828315,
the National Energy Research Scientific Computing Center (NERSC) a U.S. Department of Energy Office of Science User Facility operated under Contract No. DE-AC02-05CH11231,
the Extreme Science and Engineering Discovery Environment (XSEDE) which is supported by National Science Foundation Grant No. ACI-1548562 \cite{xsede}.
%\end{acknowledgement}

\section{AUTHOR CONTRIBUTIONS}
Y.P. established the theoretical models and supervised the project, T.S. and K.L. performed the calculations and data analysis, Y.P. and J.X. discussed the results, and all authors participated in the writing of this paper. T.S. and K.L. contributed equally to this work.

\section{ADDITIONAL INFORMATION}

%\begin{suppinfo}
\textbf{Supplementary Information}
Defect screening by spin polarized total energy calculations; defect formation energy calculations; optically detected magnetic resonance simulations for $\ti$ and $\mo$ defects; screening of SPEs by RPA calculations; single-particle levels at various levels of theory; $\mathrm{G_0W_0}$ and BSE numerical convergence and parameter setup; optical properties at RPA and BSE; nonradiative decay of $\ti$ via its PJT state; calculations of spin-orbit coupling; supercell convergence tests for nonradiative recombination lifetime.
% single-particle diagram of the $\mo$ defect in $h$-BN; $\gwrpa$ spectra for defective and pristine $h$-BN; supercell convergence of nonradiative recombination lifetime; exciton wavefunctions; 1D vs full-phonon HR factor comparison; details of spin-orbit coupling constants.

\medskip

\noindent \textbf{Competing interests} The authors declare no competing interests.

% \bibliographystyle{achemso}
% \bibliography{ref}
%%TC:endignore

\providecommand{\latin}[1]{#1}
\makeatletter
\providecommand{\doi}
  {\begingroup\let\do\@makeother\dospecials
  \catcode`\{=1 \catcode`\}=2 \doi@aux}
\providecommand{\doi@aux}[1]{\endgroup\texttt{#1}}
\makeatother
\providecommand*\mcitethebibliography{\thebibliography}
\csname @ifundefined\endcsname{endmcitethebibliography}
  {\let\endmcitethebibliography\endthebibliography}{}

\end{document}